\documentclass[aps, showpacs, showkeys,nofootinbib,floatfix ,twocolumn]{revtex4}

\usepackage{amssymb}
\usepackage{amsmath}
\usepackage{graphicx}

\begin{document}

\title{Spherical top-hat Collapse of a Viscous Unified Dark Fluid}

\author{Wei Li$^{1,2}$\footnote{corresponding author: liweizhd@126.com}}
\author{Lixin Xu$^{2}$}

\affiliation{$^{1}$Department of Physics, Bohai University, Jinzhou,
121013, China}
\affiliation{$^{2}$Institute of Theoretical Physics, Dalian
University of Technology, Dalian, 116024, P.R.China}

\begin{abstract}
In this paper, we test the spherical collapse of a viscous unified dark fluid (VUDF) which has constant adiabatic sound speed and show the nonlinear collapse for VUDF, baryons, and dark matter which are important to form the large scale structure of our Universe. By varying the values of model parameters $\alpha$ and $\zeta_{0}$, we discuss their effects on the nonlinear collapse of the VUDF model, and compare its result to $\Lambda
CDM$ model. The analyzed results show that, within the spherical top-hat collapse framework, larger values of $\alpha$ and smaller values of $\zeta_{0}$ make the structure formation earlier and faster, and the other collapse curves are almost distinguished with the curve of $\Lambda
 CDM$ model if the bulk viscosity coefficient $\zeta_{0}$ is less than $10^{-3}$. 
\end{abstract}



\maketitle

\section{Introduction}

As an competitive model to explain the lately accelerated expansion of universe, a unified dark fluid (UDF) model \cite{ref:darkdeneracy,ref:Bruni,ref:darkdeneracyxu,GCG,GCGpapers1,GCGpapers2,GCGxu}\cite{ xuNUDF,liwei} was investigated extensively in the recent years. The striking features of the UDF model are that it combines cold dark matter and dark energy and that it behaves like the cold dark matter and the dark energy at the early epoch and the late time respectively. Furthermore, it can match the image of  the $\Lambda
CDM$ model very well on the background level. Among those UDF models, one same assumption that the medium of universe is modeled as an idealized perfect fluid were taken, which means that all components of the matter-energy in our universe are considered as a perfect fluid without viscosity. However, in the recently years, more and more cosmological observations suggest that our universe is permeated by imperfect fluid, in which the negative pressure, as was argued in \cite{A. B. Balakin}, \cite{W. Zimdahl}, an effective pressure including bulk viscosity can drive the present acceleration of universe. The first attempts at creating a viscosity theory
of relativistic fluids were executed by Eckart \cite{C. Eckart} and Landau and Lifshitz \cite{L.D. Landau} who considered only first-order deviation from equilibrium. The general form of the bulk viscosity is chosen as a time-dependent function or a density-dependent function. In some literature, a density-dependent viscosity $\zeta=\zeta_{0}\rho^{m}$ coefficient is widely investigated, where the condition $\zeta_{0}>0$ ensures a positive entropy production in conformity with the second law of
thermodynamics. For simplification, we will devote ourselves to studying  the case $m=\frac{1}{2}$ , which the similar form being taken in the Ref. \cite{liwei}\cite{lixinzhou1,lixinzhou2,lixinzhou3}

For any proposed cosmological model, if it's cosmic observations cannot coincide with the theoretical calculation, it would be ruled out, so does the VUDF model. As the large-scale structure formation originates from the primordial quantum perturbations of our universe, the non-linear stages of perturbations become very important during one investigating the evolutions of density perturbations of VUDF model. A fully nonlinear analysis is a cumbersome task usually handled by hydrodynamical/N-body numerical codes (see
e.g. \cite{A.V. Maccio,N. Aghanim, M. Baldi, B. Li}). However, to best of our knowledge, the hydrodynamical/N-body numerical simulation is very implicated and expensive.

In this paper we focus on the collapse of a spherically symmetric perturbation of VUDF model, with a classical top-hat profile, instead of using the cumbersome hydrodynamical/N-body numerical simulation
. We modify the pressure pressure of UDF $p=\alpha\rho-A$ in the Ref.\cite{ref:darkdeneracyxu} into the form $p=\alpha\rho-\zeta_{0}\rho-A$ to obtain the VUDF model. As mentioned in the Ref. \cite{collapsexu}, one need to avoid the averaging problem \cite{ref:stability} when studying the non-linear perturbations. The problem comes from the fact that
\begin{equation}
\langle p\rangle=-\langle A/\rho^\beta\rangle\neq -A/\langle\rho\rangle^\beta=p(\langle\rho\rangle),
\end{equation}
in the case of $\beta\neq 0$. However, for a model with a linear relation $p=\alpha\rho-\zeta_{0}\rho-A$, it is not the problem. So it would be interesting to study the evolution of non-linear perturbation in this VUDF model because of escaping from the averaging problem.
Avoid using the hydrodynamical/$N$-body numerical simulation, we will research the large structure formation in the framework of spherical top-hat collapse for the viscous unified dark fluid (VUDF).

The paper is organized as follows. In section \ref{sec:review}, a brief introduction of the VUDF with a constant adiabatic sound speed is given. Then, we present some basic equations for spherical top-hat collapse of viscous fluid in section \ref{sec:basic}. The method and main results are summarized in section \ref{sec:perevolution}. The last section is the conclusion.

\section{Viscous Unified Dark Fluid with Constant Adiabatic Sound Speed }\label{sec:review}

In this section, we will give some basic equations of a VUDF model which has a constant
adiabatic sound speed (CASS). In order to obtain the viscous unified dark fluid, we rewrite the pressure of UDF $p=\alpha\rho-A$ in the Ref.\cite{ref:darkdeneracyxu}
into the form
\begin{eqnarray}
p_{d}&=&p-3H\zeta,
\end{eqnarray}
this expression includes the UDF model as its special case when $\zeta=0$, as for the case that $\zeta\neq0$, when adopting the normal form $\zeta=\frac{\zeta_{0}}{\sqrt{3}}\rho^{\frac{1}{2}}$ which the similar form being taken in the Ref.\cite{liwei},\cite{lixinzhou1,lixinzhou2,lixinzhou3} , we have the pressure of VUDF
\begin{eqnarray}
p_{d}=\alpha\rho_{d}-\zeta_{0}\rho_{d}-A,
\end{eqnarray}
where $A=\rho_{d0}(1+\alpha-\zeta_{0})(1-B_{s})$.
Applying the energy
conservation of VUDF, one can deduce its energy density
as the following form
\begin{eqnarray}
\rho_d&=&\rho_{d0}\left\{(1-B_s)+B_s a^{-3(1+\alpha-\zeta_{0})}\right\},
\end{eqnarray}
where the model parameters $B_s$, $\alpha$, and $\zeta_{0}$ are all in the range $[0,1]$.  So one obtains the equation of state (EoS)
\begin{eqnarray}
w_d=\frac{p_{d}}{\rho_{d}}
=\alpha-\zeta_{0}-\frac{(1+\alpha-\zeta_{0})(1-B_s)}{(1-B_s)+B_s a^{-3(1+\alpha-\zeta_{0})}},
\end{eqnarray}
and adiabatic sound speed
\begin{equation}
c_s^2=\left(\frac{\partial p_d}{\partial \rho_d}\right)_s=\frac{d p_d}{d\rho_d}=\rho_d\frac{dw_d}{d\rho_d}+w_d=\alpha-\zeta_{0},
\end{equation}
where $A$, $\zeta_{0}$ and $w_d$ are the integration constant, bulk viscosity coefficient and the equation of state (EoS) of VUDF respectively. Therefore, the Friedmann equation in a spatially flat FRW universe is given as
\begin{eqnarray}
H^{2}=H^{2}_{0}\left\{(1-\Omega_{b}-\Omega_{r})\left[(1-B_{s})+B_{s}a^{-3(1+\alpha-\zeta_{0})}\right] \right.\notag\\
\left.+\Omega_{b}a^{-3}+\Omega_{r}a^{-4}\right\},
\end{eqnarray}
where $H$ is the Hubble parameter and $H_{0}=100h\text{km s}^{-1}\text{Mpc}^{-1}$ is its present value, and $\Omega_{i}$ ($i=b,r$) are dimensionless energy density parameters, where $b$ and $r$ stand for baryon and radiation separately.

\section{Equations of Spherical Top-hat Collapse of Viscous Fluid} \label{sec:basic}

The spherical collapse (SC) as a simple analytical model was first introduced by Gunn and Gutt 1972 \cite{J. E. Gunn} in order to calculate the evolution of perturbations in falling material into a bound system which provides a way to glimpse into the nonlinear regime of perturbation theory. Usually, the SC model is used to investigate a spherically symmetric perturbation which embedded in a static, expanding or collapsing homogeneous background. In this paper we focus on the collapse of a spherically symmetric perturbation in a homogenous expanding background, with a classical top-hat profile which has the constant density \cite{ref:Fernandes2012} in the perturbed region. With the assumption of a top-hat profile, one maintains the simplified spherical collapse model as the uniformity of the perturbation throughout the collapse, which making its evolution only time-dependent. So we do not need to worry about the gradients through the collapse.

In the spherical top-hat collapse (SC-TH) model, the background evolution equations are still in the following forms
\begin{eqnarray}
\dot{\rho}&=&-3H(\rho+p),\\
 \frac{\ddot{a}}{a}&=&-\frac{4\pi G}{3}\sum_i(\rho_i+3p_i),
\end{eqnarray}
where $H=\dot{a}/a$ is the Hubble parameter.
For the perturbed region, the basic equations which depend on local quantities can be written as
\begin{eqnarray}
\dot{\rho}_c&=&-3 h(\rho_c+p_c),\\
 \frac{\ddot{r}}{r}&=&-\frac{4\pi G}{3}\sum_i(\rho_{c_i}+3p_{c_i}).
\end{eqnarray}
Here the perturbed quantities $\rho_c$ and $p_c$ are defined as $\rho_c=\rho+\delta\rho$, $p_c=p+\delta
p$; and $h=\dot{r}/r$ and $r$ are the local expansion rate and the local scale factor respectively, and furthermore $h$ relates to local expansion rate in the STHC model by \cite{ref:Abramo2009, R. A. A. Fernandes}
\begin{equation}
h=H+\frac{\theta}{3a},
\end{equation}
where $\theta\equiv\nabla\cdot \overrightarrow{v}$ is the divergence of the peculiar velocity.

So, the equations of density contrast $\delta_i=(\delta\rho/\rho)_i$ and $\theta$ are: \cite{ref:Fernandes2012,ref:Abramo2009}
\begin{eqnarray}
\dot{\delta}_i&=&-3H(c^2_{e_i}-w_i)\delta_i-[1+w_i+(1+c^2_{e_i})\delta_i]\frac{\theta}{a},\label{eq:deltat}\\
\dot{\theta}&=&-H\theta-\frac{\theta^2}{3a}-4\pi Ga\sum_i\rho_i\delta_i(1+3c^2_{e_i}),\label{eq:thetat}
\end{eqnarray}
where the effective sound speed is $c^2_{e_i}=(\delta p/\delta \rho)_i$, where $i$ stands for different energy component. The Eq. (\ref{eq:deltat}) and Eq. (\ref{eq:thetat}) can be rewritten into the form in regard to the scale factor $a$
\begin{eqnarray}
\delta'_i&=&-\frac{3}{a}(c^2_{e_i}-w_i)\delta_i-[1+w_i+(1+c^2_{e_i})\delta_i]\frac{\theta}{a^2H},\label{eq:deltaa}\\
\theta'&=&-\frac{\theta}{a}-\frac{\theta^2}{3a^2H}-\frac{3H}{2}\sum_i\Omega_i\delta_i(1+3c^2_{e_i}),\label{eq:thetaa}
\end{eqnarray}
where we have used the definition $\Omega_i=8\pi G\rho_i/3H^2$.

From the above equations, one can find that the $w_c$ and $c^2_e$ are important quantities. The definition of the EoS $w_c$ \cite{ref:Fernandes2012} is:
\begin{equation}
w_c=\frac{p+\delta p}{\rho+\delta\rho}=\frac{w}{1+\delta}+c^2_e\frac{\delta}{1+\delta}.
\end{equation}
The effective sound speed $c^2_e$ of the CASS model is given as
\begin{equation}
c^2_{e}=\frac{\delta p}{\delta\rho}=\frac{p_c-p}{\rho_c-\rho}.
\end{equation}
So, substituting the relation $p=\alpha\rho-\zeta_{0}\rho-A$ into the above equation, one has
\begin{equation}
c^2_{e}=\frac{[(\alpha-\zeta_{0})\rho_c]-A-[(\alpha-\zeta_{0})\rho-A]}{\rho_c-\rho}=\alpha-\zeta_{0}
\end{equation}.

\section{The Method and Results} \label{sec:perevolution}

In this section, we will use the spherical collapse model to investigate the non-linear evolution of the VUDF perturbations. As the baryon and VUDF are the possible components that formating the large scale structure, we will firstly consider this two components, where the results of some model parameters are come from the Ref.\cite{ref:darkdeneracyxu}: $\Omega_{d}=0.956$, $H_0=71.341 \text{km s}^{-1}\text{Mpc}^{-1}$,  and $\Omega_{b}=0.044$. With the aid of using the software {\bf Mathematica} and setting the initial conditions (ICs) $\delta_{d}$ and $\delta_{b}$ at the redshift $z=1000$ in Ref. \cite{ref:Fernandes2012}, we solve the differential equations of perturbations.

In order to explore the influences of $\alpha$ on the spherical collapse of baryon and unified dark fluid, we immobilize $\delta_{d}(z=1000)=3.5\times 10^{-3}$, $\delta_{b}(z=1000)=10^{-5}$, $\zeta_{0}=0$, and $B_s= 0.229$, but change the model parameter $\alpha=0, 10^{-3}, 10^{-2}$, and $10^{-1}$ respectively. We obtain the same calculated results as  Table \ref{tab:alpha} in the Ref. \cite{xulixin}, where the redshift $z_{ta}$ is on behalf of the turnaround redshift when the collapse is beginning. From Table \ref{tab:alpha}, one can conclude that the perturbations collapse earlier and faster for the larger values of $\alpha$ and larger values of $c^2_{e}=\alpha$.

\begin{table}[tbh]
\begin{center}
\begin{tabular}{cccc}
\hline\hline Model & $\alpha$ & $z_{ta}$ & $\delta_b(z_{ta})/\delta_d(z_{ta})$ \\ \hline
a & $0$ & $0.0678$ & $ 1.240$\\
b & $10^{-3}$ & $0.111$ & $1.211$\\
c & $10^{-2}$ & $0.138$ & $0.689$\\
d & $10^{-1}$ & $0.940$ & $0.010$ \\
\hline\hline
\end{tabular}
\caption{Models for the STHC model, where the values of $\alpha$ are small positive values due to the constraint from background evolution history. The redshift $z_{ta}$ denotes the turnaround redshift when the perturbed region begin to collapse.}\label{tab:alpha}
\end{center}
\end{table}

\begin{center}
\begin{figure}[htb]
\includegraphics[width=8cm]{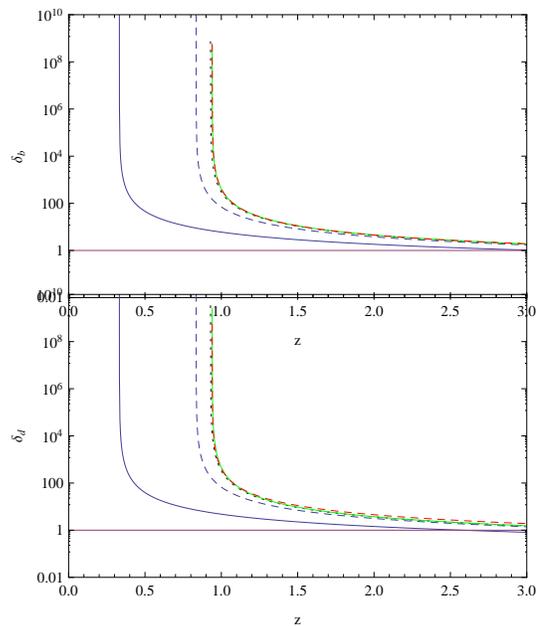}
\caption{The evolutions of density perturbations with respect to the redshift for the models $\alpha=10^{-1}$. The top and bottom panels are for baryons and VUDF respectively. Where the solid, dashed, dotted and green curved lines are for the models $\zeta_{0}=10^{-1}, 10^{-2}, 10^{-3}, 0$ respectively, beyond that the red dashed curve stands for the $\Lambda
CDM$ model. The horizon line denotes the limit of linear perturbation, i.e. $\delta=1$. The vertical parts of the curved lines denote the collapse of the perturbed regions.}\label{fig:01}
\end{figure}
\end{center}

\begin{center}
\begin{figure}[htb1]
\includegraphics[width=8cm]{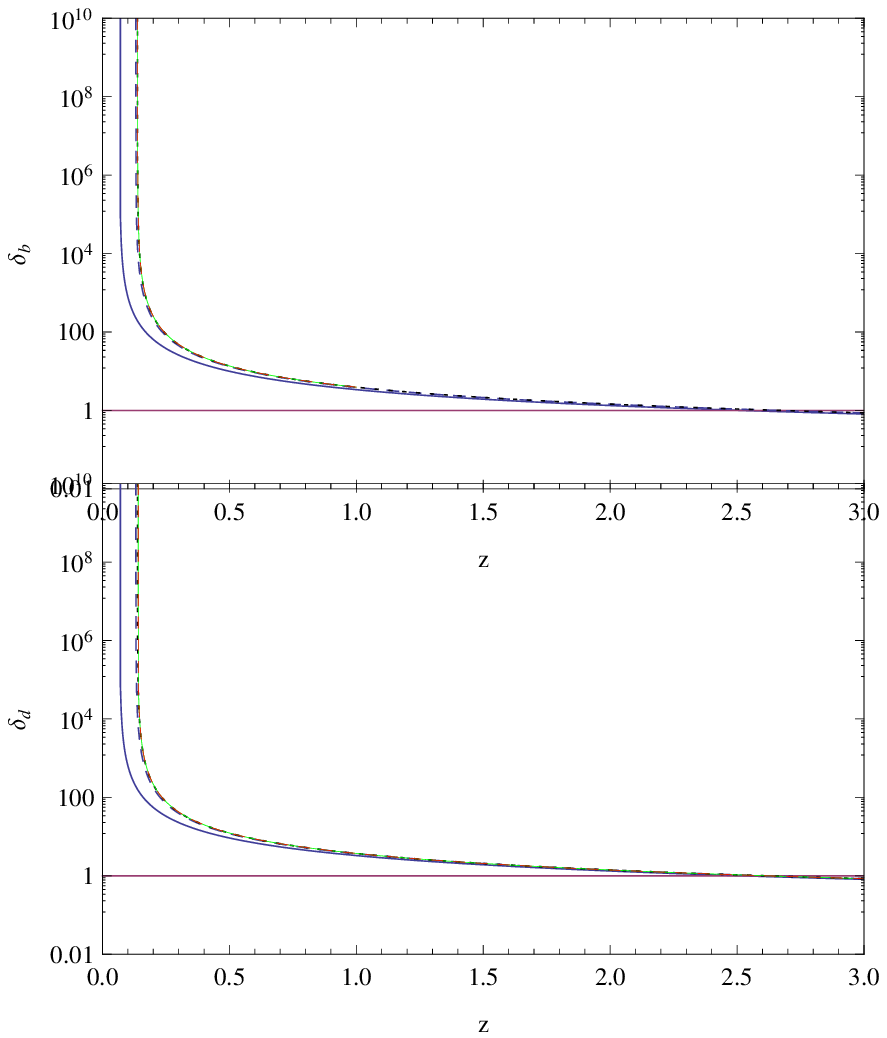}
\caption{The evolutions of density perturbations with respect to the redshift for the models $\alpha=10^{-2}$. The top and bottom panels are for baryons and VUDF respectively. Where the solid, dashed, dotted and green curved lines are for the models $\zeta_{0}=10^{-2}, 10^{-3}, 10^{-4}, 0$ respectively, beyond that the red dashed curve stands for the $\Lambda
CDM$ model. The horizon line denotes the limit of linear perturbation, i.e. $\delta=1$. The vertical parts of the curved lines denote the collapse of the perturbed regions.}\label{fig:001}
\end{figure}
\end{center}

\begin{center}
\begin{figure}[htb2]
\includegraphics[width=8cm]{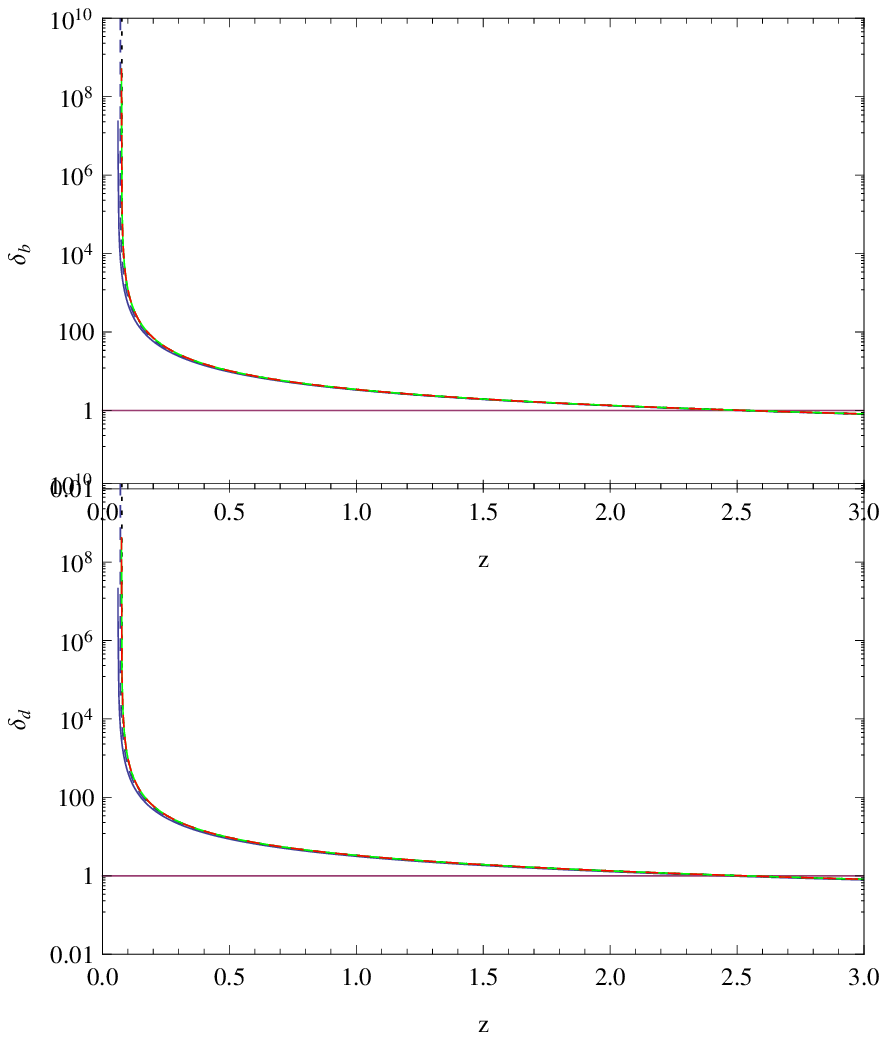}
\caption{The evolutions of density perturbations with respect to the redshift for the models $\alpha=10^{-3}$. The top and bottom panels are for baryons and VUDF respectively. Where the solid, dashed, dotted and green curved lines are for the models $\zeta_{0}=10^{-2}, 10^{-3}, 10^{-4}, 0$ respectively, beyond that the red dashed curve stands for the $\Lambda
CDM$ model. The horizon line denotes the limit of linear perturbation, i.e. $\delta=1$. The vertical parts of the curved lines denote the collapse of the perturbed regions.}\label{fig:0001}
\end{figure}
\end{center}

\begin{center}
\begin{figure}[htb3]
\includegraphics[width=8cm]{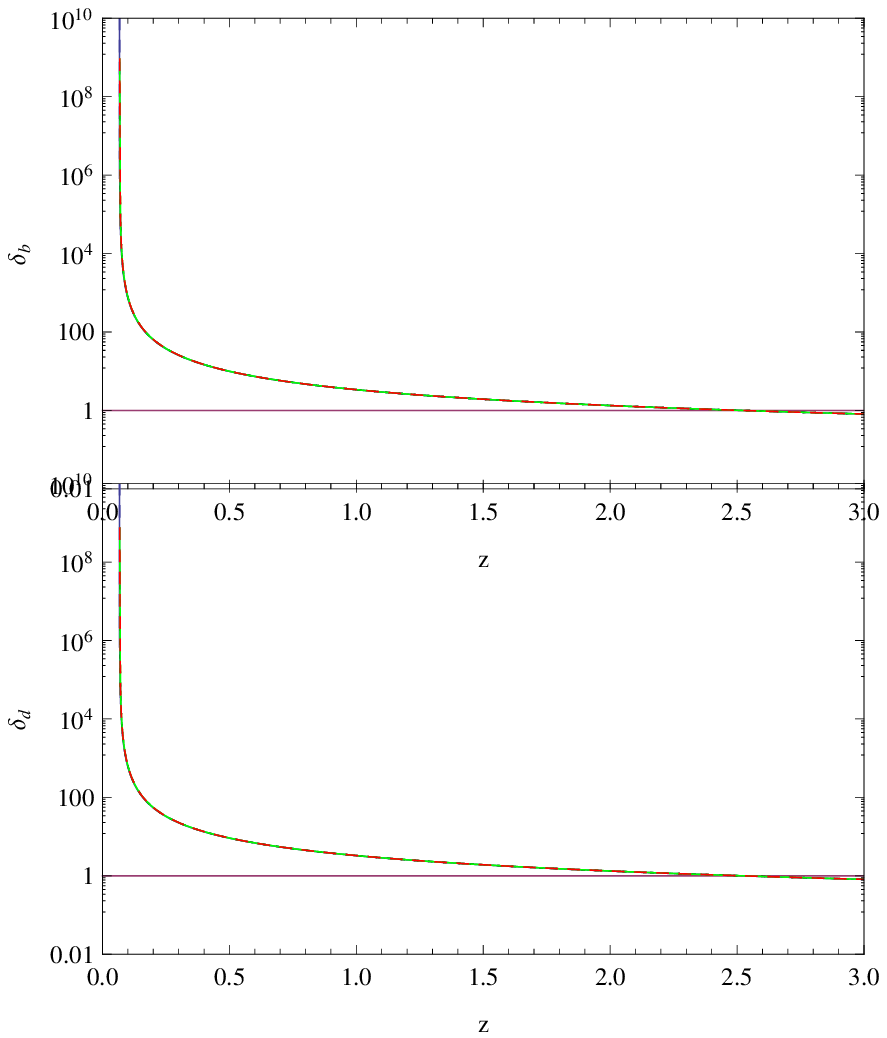}
\caption{The evolutions of density perturbations with respect to the redshift for the models $\alpha=0$. The top and bottom panels are for baryons and VUDF respectively. Where the solid, dashed, dotted and green curved lines are for the models $\zeta_{0}=10^{-3}, 10^{-4}, 10^{-5}, 0$ respectively, beyond that the red dashed curve stands for the $\Lambda
CDM$ model. The horizon line denotes the limit of linear perturbation, i.e. $\delta=1$. The vertical parts of the curved lines denote the collapse of the perturbed regions.}\label{fig:0}
\end{figure}
\end{center}

\begin{center}
\begin{figure}[htb3]
\includegraphics[width=8cm]{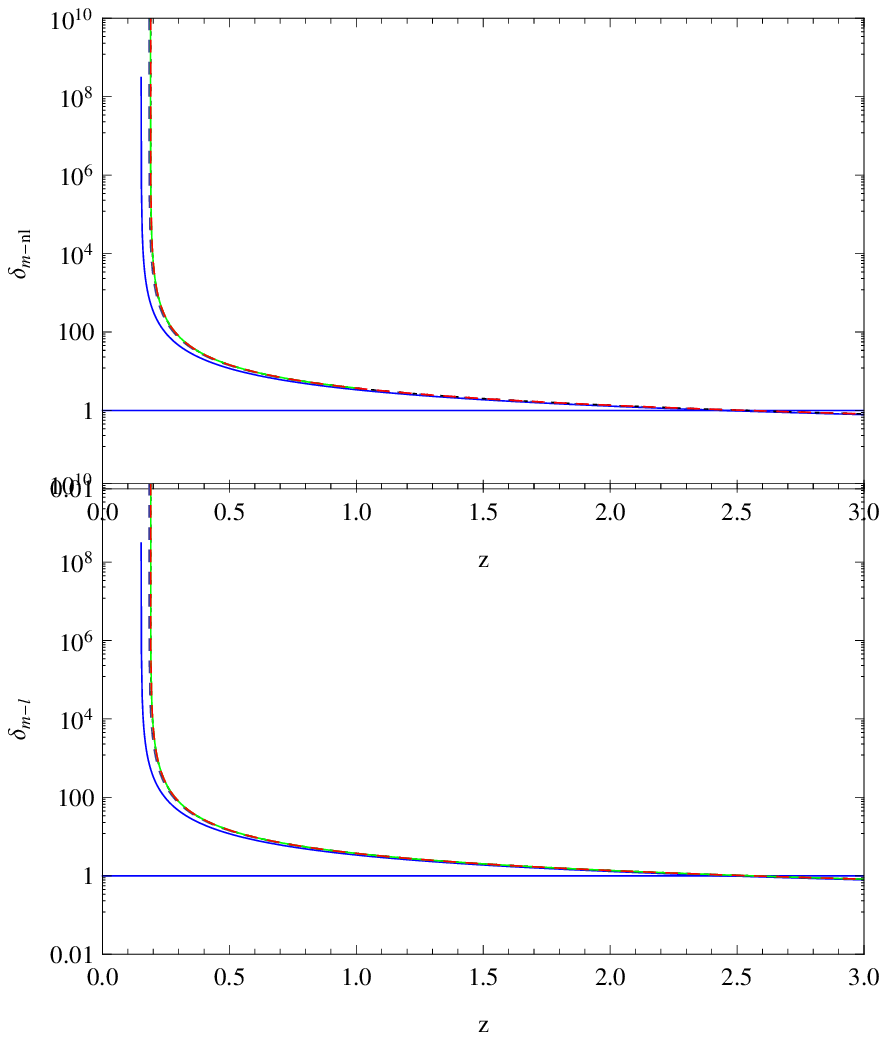}
\caption{The evolutions of density perturbations with respect to the redshift for the VUDF model and $\Lambda
CDM$ model. The top and bottom panels are for the nonlinear matter perturbation and the linear matter perturbation respectively. Where the solid, dashed, dotted and green curved lines are for the models $\zeta_{0}=10^{-2}, 10^{-3}, 10^{-4}, 0$, beyond that the red dashed curve stands for the $\Lambda
CDM$ model. The horizon line denotes the limit of linear perturbation, i.e. $\delta=1$. The vertical parts of the curved lines denote the collapse of the perturbed regions.}\label{fig:m}
\end{figure}
\end{center}

\begin{center}
\begin{figure}[htb]
\includegraphics[width=8cm]{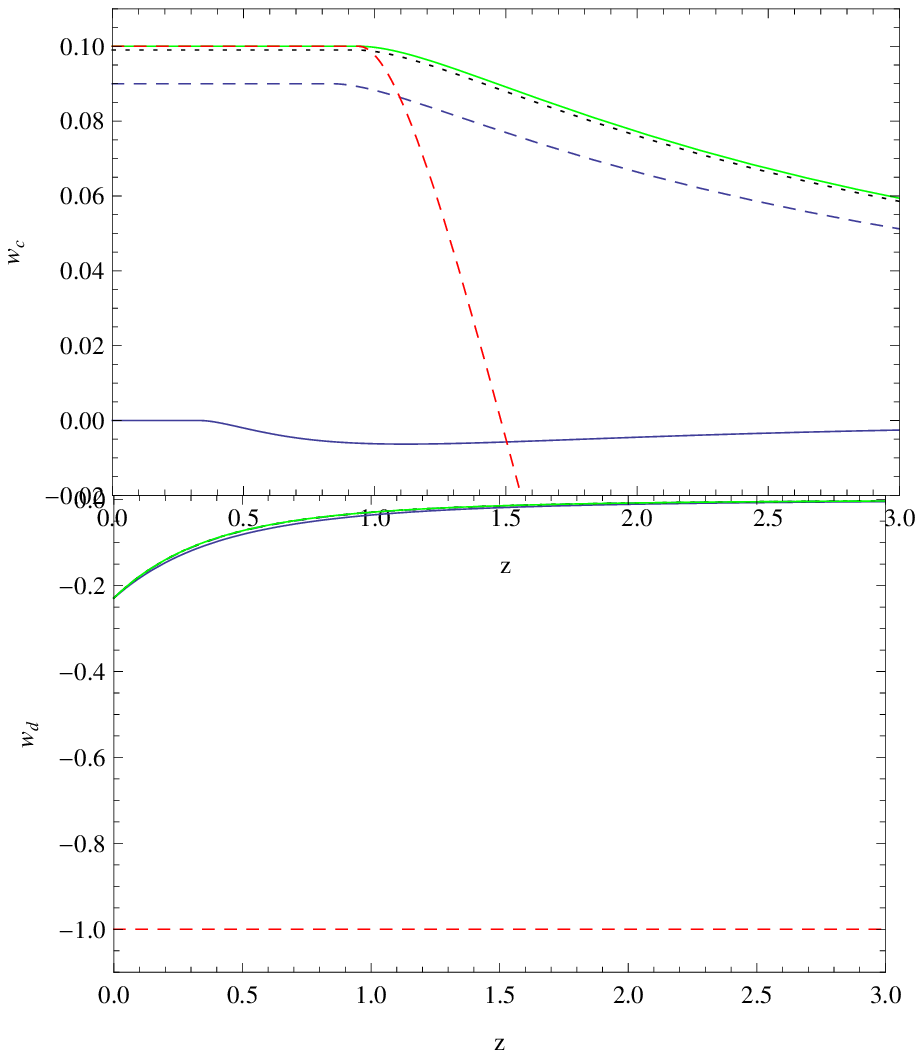}
\caption{The evolutions for equation of state $w$ with respect to the redshift $z$ for different models $\alpha=10^{-1}$. The top and bottom panels are for $w_c$ and $w_d$ respectively. Where the solid, dashed, dotted and green curved lines are for $\zeta_{0}=10^{-1}, 10^{-2}, 10^{-3}, 0$ respectively, beyond that the red dashed curve stands for the $\Lambda
CDM$ model.}\label{fig:w01}
\end{figure}
\end{center}

\begin{center}
\begin{figure}[htb]
\includegraphics[width=8cm]{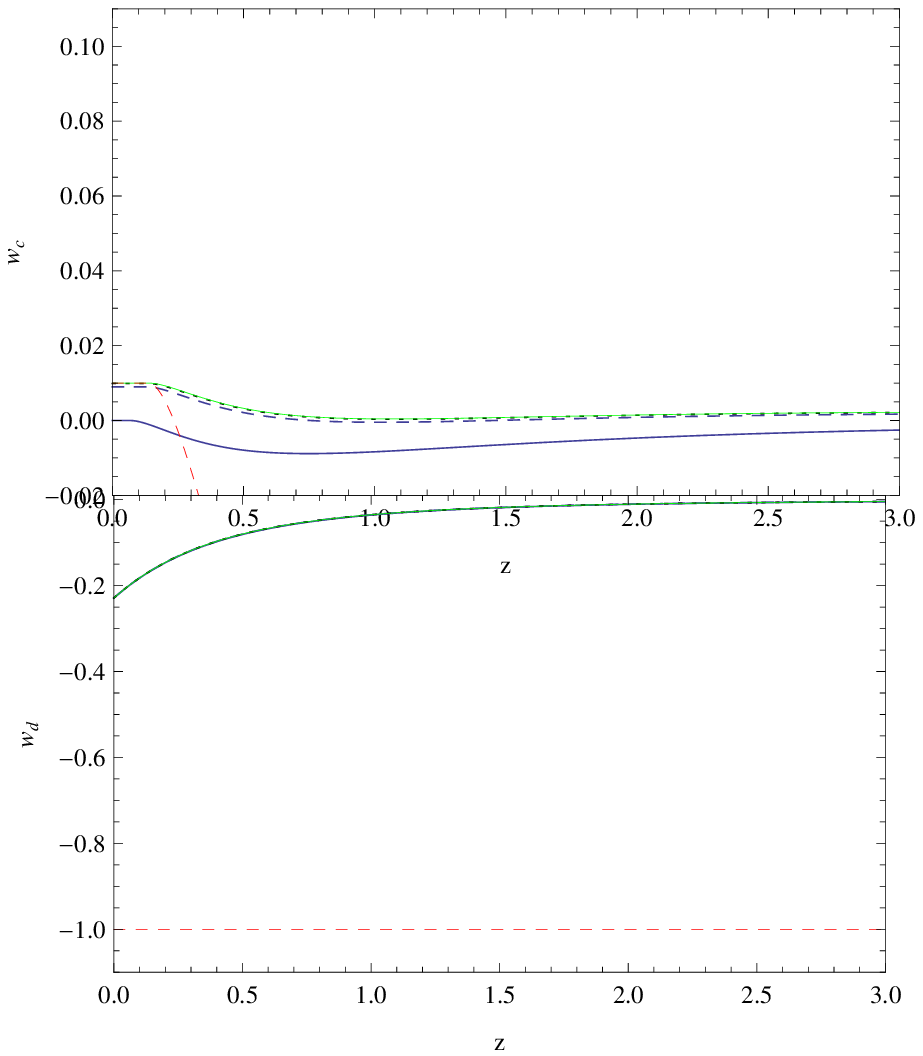}
\caption{The evolutions for equation of state with respect to the redshift $z$ for different models $\alpha=10^{-2}$. The top and bottom panels are for $w_c$ and $w_d$ respectively. Where the solid, dashed, dotted and green curved lines are for $\zeta_{0}=10^{-2}, 10^{-3}, 10^{-4}, 0$ respectively, beyond that the red dashed curve stands for the $\Lambda
CDM$ model.}\label{fig:w001}
\end{figure}
\end{center}

\begin{center}
\begin{figure}[htb]
\includegraphics[width=8cm]{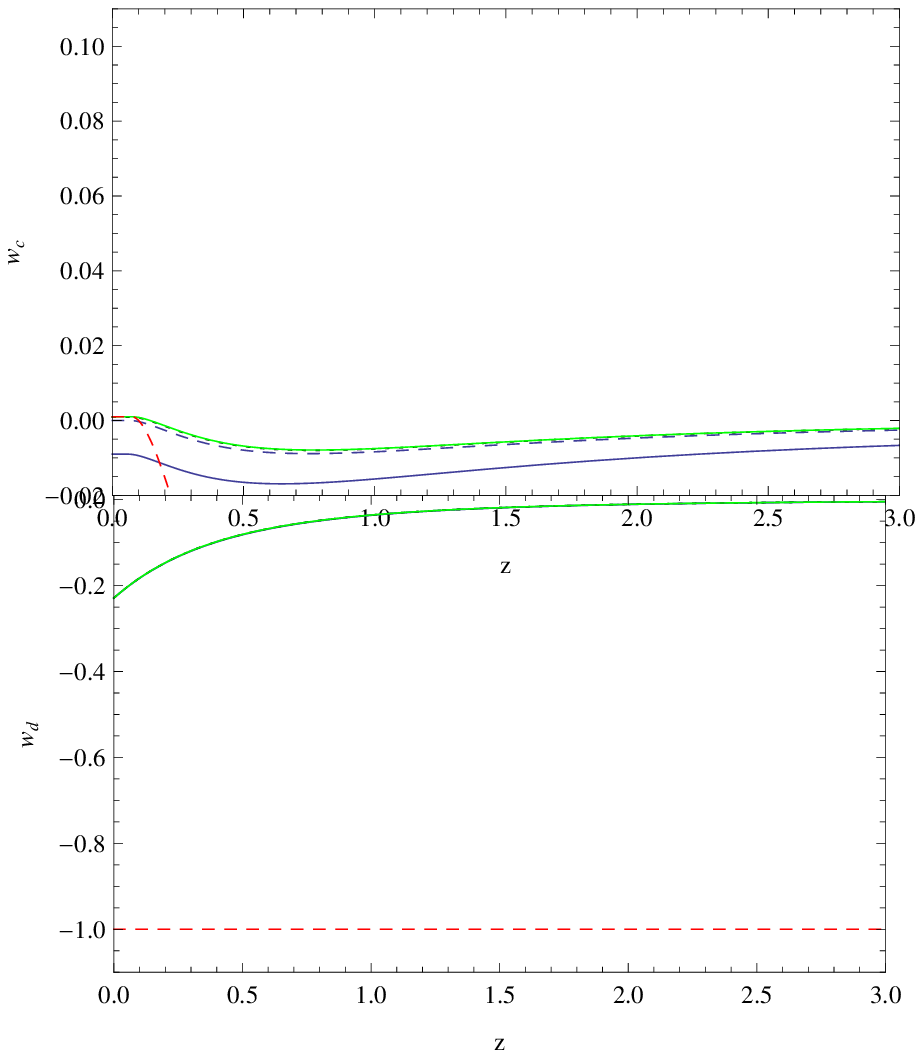}
\caption{The evolutions for equation of state with respect to the redshift $z$ for different models $\alpha=10^{-3}$. The top and bottom panels are for $w_c$ and $w_d$ respectively. Where the solid, dashed, dotted and green curved lines are for $\zeta_{0}=10^{-2}, 10^{-3}, 10^{-4}, 0$ respectively, beyond that the red dashed curve stands for the $\Lambda
CDM$ model.}\label{fig:w0001}
\end{figure}
\end{center}

\begin{center}
\begin{figure}[htb]
\includegraphics[width=8cm]{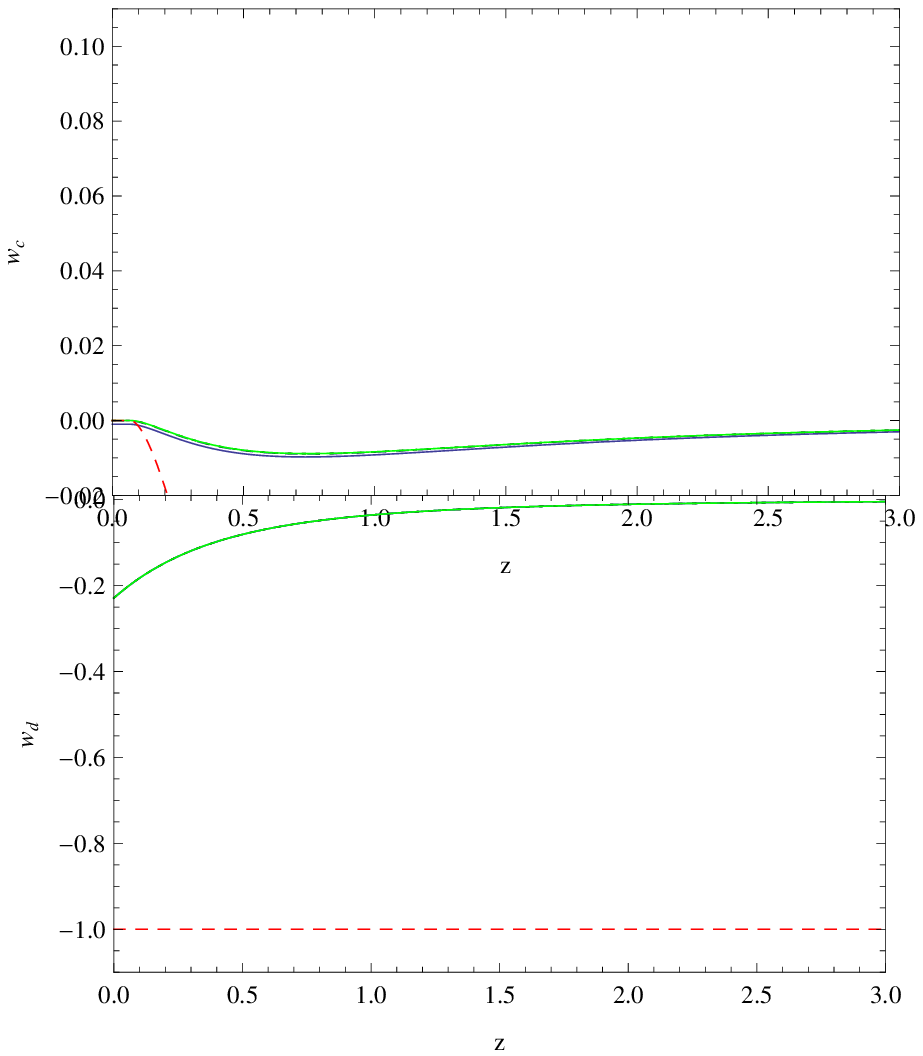}
\caption{The evolutions for equation of state with respect to the redshift $z$ for different models $\alpha=0$. The top and bottom panels are for $w_c$ and $w_d$ respectively. Where the solid, dashed, dotted and green curved lines are for $\zeta_{0}=10^{-3}, 10^{-4}, 10^{-5}, 0$ respectively, beyond that the red dashed curve stands for the $\Lambda
CDM$ model.}\label{fig:w0}
\end{figure}
\end{center}

\begin{center}
\begin{figure}[htb]
\includegraphics[width=8cm]{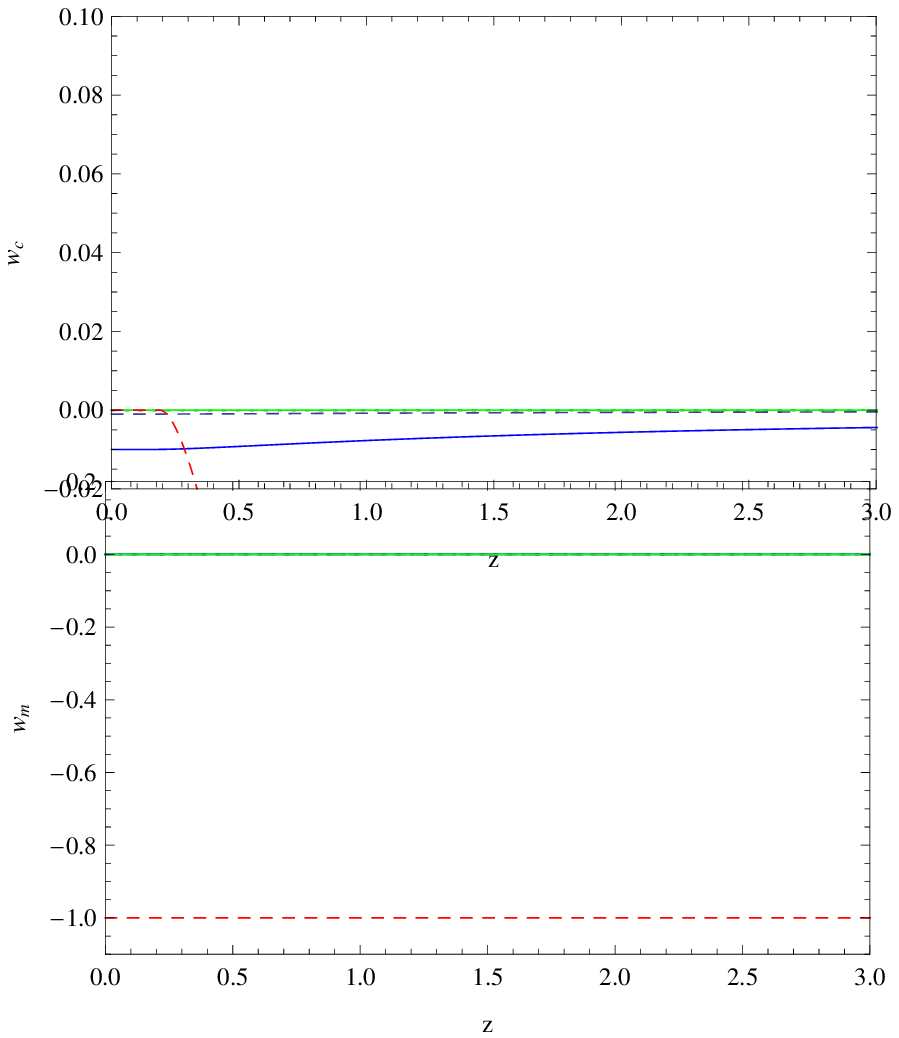}
\caption{The evolutions for equation of state $w$ with respect to the redshift $z$ for the VUDF model and $\Lambda
CDM$ model. The top and bottom panels are for $w_c$ and $w_m$ respectively. Where the solid, dashed, dotted and green curved lines are for the models $\zeta_{0}=10^{-2}, 10^{-3}, 10^{-4}, 0$, beyond that the red dashed curve stands for the $\Lambda
CDM$ model.}\label{fig:mw}
\end{figure}
\end{center}

In the following, we will show the influence of bulk viscosity coefficient $\zeta_{0}$ on the evolution of the density perturbations of baryon and VUDF. Here we alter the values of $\zeta_{0}$ for the different models which corresponding to $\alpha=10^{-1}, 10^{-2}, 10^{-3},$ and $0$ respectively. For the viscous unified dark fluid model, $\Lambda
CDM$ model is recovered when model parameters $\zeta_{0}=0$ and $B_{s}=0$ are taken. In order to compare the VUDF model with $\Lambda
 CDM$ model, we plot collapse curves of these two models in the same figure. The corresponding evolutions of density perturbations are shown in Figure \ref{fig:01}, \ref{fig:001}, \ref{fig:0001}, \ref{fig:0}. Moreover,   dark matter is regained if model parameters $B_{s}=1$ and $\alpha=0$ are respected, so we plot the non-linear and linear perturbations evolution curves of dark matter in Figure \ref{fig:m}. From the first four figures, one can see that the smaller value of $\alpha$ is taken, the more unconspicuous influence on the collapse is gained, for example in the Figure \ref{fig:0}, when $\alpha=0$, one almost unable to distinguish that five curves. Seeing from the Figure \ref{fig:01}-\ref{fig:m}, the horizon line denotes the limit of linear perturbation, i.e. $\delta=1$ and the vertical parts of the curved lines denote the collapse of the perturbed regions, therefore, one can see that smaller value of bulk viscosity coefficient $\zeta_{0}$ can result in earlier collapse, that is to say, larger value of bulk viscosity coefficient $\zeta_{0}$ can make the collapse more later, these results are compatible with the well-known convention that the value of bulk viscosity coefficient $\zeta_{0}$ should be not too large. Furthermore, one can find that when the bulk viscosity coefficient $\zeta_{0}$ is less than $10^{-3}$, the other collapse curves almost overlap together with the curve of $\Lambda
 CDM$ model.
It's time to show the influence of $\zeta_{0}$ on the evolution of the equation
of state (EoS) of the VUDF $w_{d}$ and the EoS of the collapse region $w_{c}$. Observing the evolving curves of $w_{c}$ in the Figure \ref{fig:w01}-\ref{fig:w0}, one can easily conclude that higher values of $\zeta_{0}$ result in values of $w_{c}$ closer
and higher to $w_{c}=0$ during the collapse as shown in Figure \ref{fig:w01} and \ref{fig:w001}, and result in values of $w_{c}$ closer
and lower to $w_{c}=0$ during the collapse as shown in Figure \ref{fig:w0001} and \ref{fig:mw}, but result in values of $w_{c}$ almost overlapping together as as shown in Figure \ref{fig:w0}. However, the effects of $\zeta_{0}$ on the evolution of the equation $w_{d}$ is very different comparing to the results above. Apart from the almost distinguishable influence of $\zeta_{0}$ on $w_{d}$ as shown in Figure \ref{fig:w001}-\ref{fig:w0}, we know that smaller $\zeta_{0}$ make the curves of $w_{d}$ higher as shown in Figure \ref{fig:w01}. Base on the discussion above, we can draw a conclusion that the influence of $\zeta_{0}$ on the evolution of  $w_{d}$ and $w_{c}$ are enhanced as increasing the values of $\alpha$. Apart that, from Figure \ref{fig:w01}-\ref{fig:mw}, we can easily conclude that, whatever the value of parameters $\alpha$ and $\zeta_{0}$ are, the evolution curves of $w_{d}$ for the VUDF model and $\Lambda
CDM$ model are very different. However, for the evolution curves of $w_{c}$, the curves of VUDF model(when the value of $\zeta_{0}$ is less than $10^{-3}$) and $\Lambda
CDM$ model are match together at the late time, and that the value of $\alpha$ is larger, the superposition is happened earlier.

Through the calculation and analysis above, for the VUDF model, it is possible to format the large scale structure. Also, it is obvious that the model parameters $\zeta_{0}$ and $\alpha$ have influence on the density perturbations evolutions.

\section{Conclusion} \label{sec:conclusion}

In this paper, we investigated the density perturbations of a VUDF model with a constant adiabatic sound speed in the framework of spherical top-hat collapse, the results showed that it is possible to form large scale structure in the VUDF model. We studied their influence on the evolutions of perturbation through varying the values of $\zeta_{0}$ and $\alpha$. Through the calculation and analysis, we concluded that smaller values of $\zeta_{0}$ and larger values of $\alpha$ can make the density perturbations collapse earlier and faster, and that the other collapse curves almost overlap together with the curve of $\Lambda
 CDM$ if the bulk viscosity coefficient $\zeta_{0}$ is less than $10^{-3}$. Furthermore, we can also conclude that the influence of $\zeta_{0}$ on the evolution of  $w_{d}$ and $w_{c}$ are enhanced as increasing the values of $\alpha$. In the following work, we will try to apply  the spherical collapse to other cosmological models and compare the simulated results with the observed large scale structure of universe.

\section{Acknowledgements}
L. Xu's work is supported in part by NSFC under the Grants No. 11275035 and "the Fundamental Research Funds for the Central Universities" under the Grants No. DUT13LK01.

\end{document}